\documentclass[conference]{IEEEtran}
\IEEEoverridecommandlockouts
\usepackage{cite}
\usepackage{amsmath,amssymb,amsfonts}
\usepackage{algorithmic}
\usepackage{graphicx}
\usepackage{textcomp}
\usepackage{xcolor}
\usepackage{fontenc}
\usepackage[procnames]{listings}
\usepackage{lipsum,mwe,cuted}
\usepackage{float}%%%%提供浮动体的[H]选项，进而取消浮动
\usepackage{caption}%%提供captionof命令
\usepackage{booktabs} 
\usepackage{multirow}
\usepackage{tikz}
\usepackage{colortbl}
\usepackage{makecell}
\usepackage{pifont}
\usepackage{multirow}
\usepackage{mdframed}
\usepackage{diagbox}
\usepackage{booktabs}
\usepackage{array}
\usepackage[numbers, square]{natbib}
\usepackage[hidelinks, colorlinks=true, linkcolor=black, urlcolor=blue, citecolor=black]{hyperref}

\definecolor{yanse}{HTML}{FFC773} % 用HTML颜色代码定义名为"yanse"的颜色
\definecolor{mygreen}{HTML}{A4E2C6} % 用HTML颜色代码定义名为"yanse"的颜色

\def\BibTeX{{\rm B\kern-.05em{\sc i\kern-.025em b}\kern-.08em
    T\kern-.1667em\lower.7ex\hbox{E}\kern-.125emX}}
\begin{document}

\title{AutoVCoder: A Systematic Framework for Automated Verilog Code Generation using LLMs}

\author{\IEEEauthorblockN{Mingzhe Gao$^{1}$,
Jieru Zhao$^{1*}$, Zhe Lin$^{2*}$, Wenchao Ding$^{3}$, Xiaofeng Hou$^{1}$, Yu Feng$^{1}$, Chao Li$^{1}$, Minyi Guo$^{1}$}
\IEEEauthorblockA{${}^{1}$Shanghai Jiao Tong University, ${}^2$Sun Yat-sen University, ${}^3$Fudan University\\
\{a823337391z,zhao-jieru\}@sjtu.edu.cn, linzh235@mail.sysu.edu.cn, guo-my@cs.sjtu.edu.cn}\vspace{-1cm}}

\maketitle
\begingroup\renewcommand\thefootnote{*}
\footnotetext{Jieru Zhao and Zhe Lin are the corresponding authors.}
\endgroup

\setlength{\abovedisplayskip}{3pt}
\setlength{\belowdisplayskip}{3pt}

\begin{abstract}
%Large language models are adept at rapidly generating software code in languages such as C, C++, and Python, largely due to extensive datasets. However, their performance in generating hardware-related code, particularly in Verilog, has not been satisfactory. To address this issue, we develope a systematic framework that includes fine-tuning and a Retrieval-Augmented Generation (RAG) module. This framework significantly improves the model's ability to generate Verilog code and enhances the quality of its output. Our data show that this fine-tuned model not only exceeds the performance of ChatGPT-3.5 on specific benchmarks but also shows improved capability in Verilog code generation through the use of the RAG strategy.

Recently, the use of large language models (LLMs) for software code generation, e.g., C/C++ and Python, has proven a great success. However, LLMs still suffer from low syntactic and functional correctness when it comes to the generation of register-transfer level (RTL) code, such as Verilog. To address this issue, in this paper, we develop AutoVCoder, a systematic open-source framework that significantly improves the LLMs’ correctness of generating Verilog code and enhances the quality of its output at the same time. Our framework integrates three novel techniques, including a high-quality hardware dataset generation approach, a two-round LLM fine-tuning method and a domain-specific retrieval-augmented generation (RAG) mechanism. Experimental results demonstrate that AutoVCoder outperforms both industrial and academic LLMs in Verilog code generation. Specifically, AutoVCoder shows a 0.5\% and 2.2\% improvement in functional correctness on the EvalMachine and EvalHuman benchmarks compared with BetterV, and also achieves a 3.4\% increase in syntax correctness and a 3.4\% increase in functional correctness on the RTLLM benchmark compared with RTLCoder.
\end{abstract}

\section{Introduction}

Large Language Models (LLMs) has increasingly captured the attention of the academia and industry. In the realm of programming, LLMs have demonstrated remarkable success in generating software code, automating and streamlining the development process of programming languages like C, C++, and Python. Recently, some representative works~\cite{wang2021codet5,nijkamp2022codegen,zheng2023codegeex,chen2021evaluating, nijkamp2023codegen2, fried2022incoder}, including CodeT5~\cite{wang2021codet5}, CodeGen~\cite{nijkamp2022codegen}, CodeGeeX~\cite{zheng2023codegeex}, have made tremendous breakthroughs in augmenting LLMs for software code generation. Additionally, commercial tools such as Copilot~\cite{github2021copilot} and GPT-4~\cite{openai2024gpt4} have demonstrated notable performance in code generation. The progress is largely driven by advances in model architecture, training techniques, and most importantly, the vast amounts of data on which these models are trained. 

%\textcolor{red}{3 reasons -- 1. hd code is different from sw code}

%\textcolor{red}{          -- 2. hd data is limit}

%\textcolor{red}{          -- 3. hd benchmarks is limit}

%However, despite the impressive performance of LLMs in software code generation, their effectiveness in the hardware domain, particularly in generating hardware-specific codes such as Verilog, are still remains underwhelming. This difference is due to unique challenges in the hardware context. First, hardware languages often incorporate a higher level of specificity and complexity in their syntax and semantic requirements compared to general-purpose programming languages. Second, the volume of high-quality, annotated hardware language data available for training is considerably smaller than that for software languages, limiting the learning potential of models in this area. Third, compared to software code generation which has a rich set of benchmarks like HumanEval\cite{chen2021codex}, MBPP\cite{austin2021program} and APPS\cite{hendrycksapps2021}, benchmarks for hardware code generation is limited and the variety is not as extensive, which makes it difficult for us to assess the capability of LLMs in generating hardware code.

However, despite the tremendous advancements in software code generation achieved by LLMs, their effectiveness in the hardware domain, particularly in constructing hardware designs using Verilog, still remains underwhelming. This gap can be attributed to two unique challenges of hardware design. First, RTL languages often incorporate greater domain specificity and complexity in their syntax and semantics compared to the software programming languages. Second, the volume of high-quality hardware design datasets available for training is considerably smaller than that for software languages, limiting the learning capability of large models. 
%Third, compared to software code generation which has a rich set of benchmarks like HumanEval\cite{chen2021codex}, MBPP\cite{austin2021program} and APPS\cite{hendrycksapps2021}, benchmarks for hardware code generation is limited and the variety is not as extensive, which makes it difficult for us to assess the capability of LLMs in generating hardware code.
%\textcolor{red}{3 reasons -- 1. DATE 2023 benchmark, try to learn the ability by fine-tune LLM}
%\textcolor{red}{          -- 2. DAC 2023, try to learn the ability by fine-tune LLM}
%\textcolor{red}{          -- 3. More prompt engineering, many work}
%\textcolor{red}{          -- 4. ASP-DAC 2023, benchmark}
%\textcolor{red}{          -- 5. RTLCoder, new training scheme}
%\textcolor{blue}{Previous works in the domain of hardware code generation have attempted to bridge this gap. With the advent of ChatGPT-3.5 and GPT-4, numerous projects have emerged that use LLMs to generate Verilog code using prompt engineering. ChipGPT\cite{chang2023chipgpt} introduced an automatic chip generation framework through prompt engineering that allows ChatGPT-3.5 to generate circuits with lower power consumption and smaller size. RTLLM\cite{lu2024rtllm} developed a self-planning prompt engineering approach that enhances the syntactic and functional accuracy of Verilog code generated by ChatGPT-3.5 and GPT-4. RTLFixer\cite{tsai2024rtlfixer} improved syntax accuracy by using compiler-generated error messages to feedback into the LLM across multiple rounds of queries.}

Prior research in hardware code generation has attempted to bridge this gap and can be categorized into two types: prompt engineering and supervised fine-tuning. Prompt engineering improves the quality of LLMs' generated Verilog code by adjusting the descriptions and structures of prompts. Without altering model parameters, this method can be easily adopted and implemented. ChipGPT~\cite{chang2023chipgpt} introduces an automatic chip generation framework through prompt engineering, allowing ChatGPT-3.5 to generate circuits with lower power consumption and smaller size. RTLLM~\cite{lu2024rtllm} develops a self-planning prompt engineering approach that enhances the syntactic and functional accuracy of Verilog code generated by ChatGPT-3.5 and GPT-4. RTLFixer~\cite{tsai2024rtlfixer} and AutoChip~\cite{thakur2023autochip}  improves syntactic accuracy by utilizing compiler-generated error messages as feedback to LLM across multiple rounds of queries. While prompt engineering is convenient and requires less preparation than fine-tuning, it does not fundamentally improve the underlying ability of LLMs for RTL code generation, making supervised fine-tuning a necessary step. 
%Thakur et al.~\cite{date23} introduced the first fine-tuning framework with early LLMs like GPT-2~\cite{radford2019language} and CodeGen~\cite{nijkamp2022codegen}, improving the ability of LLM to generate Verilog code for problems from the HDL website~\cite{hdlbits}. 
%\textcolor{red}{the accuracy of the results is low due to the scale of the LLMs. }
%Dehaerne et al.~\cite{dehaerne2023deep} utilized CodeGen fine-tuned on GitHub code to compare the sentence similarity between the LLMs output and the ground truth verilog code, which overlook the significant differences between code generation and other NLP tasks. NVIDIA research team introduced ChipNeMo~\cite{liu2024chipnemo}, which collects internal company data to fine-tune LLaMA~\cite{touvron2023llama}, resulting in an LLM capable of generating QA pairs and Verilog code, yet it remains closed-source. RTLCoder~\cite{liu2024rtlcoder} proposed a fine-tuning approach that employs a scoring mechanism during training, which improves the effectiveness of the trained LLMs. However, their dataset is limited in size, which may not be suitable for broader tasks, and the scoring system used during training requires meticulous design.
%To summarize, \textcolor{red}{need to rewrite}
%there hasn’t been a framework that effectively enhances the entire LLM model construction process dedicated to Verilog code generation, i.e., from dataset collection to model training to prompt engineering.
Thakur et al.~\cite{date23} and Dehaerne et al.~\cite{dehaerne2023deep} adopt full fine-tuning which adjusts parameters of the entire LLM model with their dataset collected from GitHub. However, the lack of adequate data cleaning and task-specific training influences their functional accuracy. 
ChipNeMo~\cite{liu2024chipnemo} from Nvidia deploys a two-round fine-tuning process with their in-house data, while only the first round can benefit RTL code generation. The dataset and model are not released to the public. Meanwhile, its functional accuracy is not satisfying due to the same issue as~\cite{date23,dehaerne2023deep}.
% on the VerilogEval benchmark ~\cite{liu2023verilogeval}.
% introduced the first fine-tuning framework with early LLMs like GPT-2~\cite{radford2019language} and CodeGen~\cite{nijkamp2022codegen}, improving the ability of LLM to generate Verilog code for problems from the HDL website~\cite{hdlbits}. 
% However, their dataset lacks quality filter mechanisms and task-specific training, resulting in low functionality accuracy of their models on existing benchmarks. 
% Dehaerne et al.~\cite{dehaerne2023deep} utilized a similar full fine-tuning method to compare the sentence similarity between the LLMs output and the ground truth verilog code, which induce the same issue as \cite{date23}. 
To improve the model performance, 
% fine tune models on datasets with higher quality,
VerilogEval~\cite{liu2023verilogeval} and the most recent work, RTLCoder~\cite{liu2024rtlcoder}, utilize ChatGPT-3.5 to generate high-quality problem-code pairs as the dataset for the single-round task-specific fine-tuning, demonstrating relatively good results on existing benchmarks. However, their small synthetic dataset lacks diversity, influencing the generality and making it hard to achieve higher accuracy. BetterV~\cite{betterv} simultaneously fine-tunes the model and trains a generative discriminator to assist in Verilog code generation, which increases the difficulty of deployment.

%NVIDIA research team utilizes the low rank adaptation (LoRA) fine-tuning method which tunes less parameters and is more efficient.

To summarize, previous methods fail to balance the diversity and quality of dataset. Additionally, the single-round domain-adapted fine-tuning limits the ability of LLM to maintain high code quality while generating diverse outputs. Meanwhile, LLM faces the hallucination problem and may incorporate software coding habits into Verilog code generation, potentially causing syntactic and functional errors. However, few of prior methods notices this issue.

%\textcolor{blue}{While prompt engineering has been somewhat effective, it doesn't fundamentally alter the underlying ability of LLMs, making fine-tuning a necessary step. \cite{date23} introduces the first fine-tuning framework using early large language models like GPT-2\cite{radford2019language} and CodeGen, exploring the ability to generate Verilog code problems from the HDL website\cite{hdlbits}. \cite{dehaerne2023deep} utilized CodeGen fine-tuned on GitHub code to compare the sentence similarity between the LLMs output and the ground truth verilog code. }

%\textcolor{red}{we? a systematic framework for data collect, data filter and pre-process}

In this paper, we propose AutoVCoder, a systematic open-source framework that strengthens the capability of LLMs to automatically generate high-quality Verilog code. AutoVcoder enhances LLMs to generate syntactically and functionally correct Verilog code, addressing the gap to apply LLMs for hardware code generation. Our key contributions are summarized as follows: 

\begin{enumerate}
\item We propose an effective and automated dataset generation approach that generates high-quality and diverse RTL code samples. 
% that equips LLMs with the ability to produce Verilog-based hardware designs with minimal manual intervention.
\item We introduce a two-round LLM fine-tuning method to improve the ability of LLMs for Verilog code generation. %with a dataset quality scoring mechanism and an automatic verification flow \textcolor{red}{scoring and verification is for dataset generation? not fine-tuning? should demonstrate characteristics of fine-tuning}
%\item We present a retriever-augmented generation(RAG) method that provides more suitable guidance prompts for posed questions and integrates them effectively, and enhances the scalability of LLM.
\item We present a domain-specific retrieval-augmented generation (RAG) module that provides more constructive prompts to further enhance the syntactic and functional correctness of generated RTL designs.
\end{enumerate}

Experimental results demonstrate that AutoVCoder outperforms both industrial and academic LLMs in Verilog code generation. Specifically, AutoVCoder shows a 0.5\% and 2.2\% improvement in functional correctness on the EvalMachine and EvalHuman benchmarks compared with BetterV, and also achieves a 3.4\% increase in syntax correctness and a 3.4\% increase in functional correctness on the RTLLM benchmark compared with RTLCoder.

% Experimental results show that AutoVCoder outperforms both industrial and academic LLMs in Verilog code generation, with an improvement of 3.4\% in syntax correctness and 3.4\% in functional correctness, compared to the state-of-the-art method, RTLCoder \cite{liu2024rtlcoder}.

%\begingroup\renewcommand\thefootnote{*}
%\footnotetext{Result from GPT-4.}

\section{Preliminaries}

\subsection{Large Language Model for Code Generation}

%Large Language Models(LLMs) have revolutionized the field of natural language processing(NLP). Their ability to generate coherent and contextually relevant text has made them particularly useful for a wide range of applications, including code generation. In the context of software development, LLMs are trained on vast corpora of code across various programming languages, enabling them to learn syntax, semantics, and some level of logical flow inherent to software coding. For code generation, LLMs are often employed in an autoregressive manner, where the model predicts the next token in a sequence given the previous tokens. This capability allows LLMs to write entire blocks of code based on a prompt that specifies the desired functionality. The effectiveness of these models in generating syntactically correct and logically coherent code snippets has been demonstrated in various studies\cite{wang2021codet5, nijkamp2022codegen, zheng2023codegeex, chen2021evaluating, wang2023codet5plus, le2022coderl, nijkamp2023codegen2, fried2022incoder, github2021copilot, openai2024gpt4}, showcasing their potential to automate certain aspects of software development and reduce the cognitive load on human developers.

Large Language Models (LLMs) have revolutionized the field of natural language processing (NLP). Their ability to generate coherent and contextually relevant text has made them particularly useful for various applications such as code generation. In the context of software code generation, LLMs are trained on a vast amount of code across various programming languages, enabling them to learn syntax, semantics, and some level of logical flow inherent to software coding. Then LLMs are employed in an autoregressive manner, predicting the next token in a sequence given previous tokens. This capability allows LLMs to generate entire blocks of code based on a prompt that specifies the desired functionality. The effectiveness of these models in generating syntactically correct and logically coherent code snippets has been demonstrated in previous studies~\cite{wang2021codet5, nijkamp2022codegen, zheng2023codegeex, chen2021evaluating, nijkamp2023codegen2, fried2022incoder, github2021copilot, openai2024gpt4}, showcasing their potential to automate software development and reduce workload of developers.

However, the transition of LLMs from generating software code to generating RTL code like Verilog presents unique challenges. It requires a deep understanding of domain knowledge, such as hardware architecture, circuit design and low-level constraints, to generate a high-quality hardware design. This can be hard to achieve with standard LLM training datasets. Therefore, while previous works~\cite{lu2024rtllm, tsai2024rtlfixer, date23, liu2024rtlcoder} show that powerful LLMs like GPT-3.5 can handle HDL syntax due to similarities between programming languages, their performance in generating functionally correct hardware designs is still unsatisfying without additional domain-specific processing and fine-tuning.

\begin{figure}
        \centering
	 \includegraphics[width=0.95\linewidth]{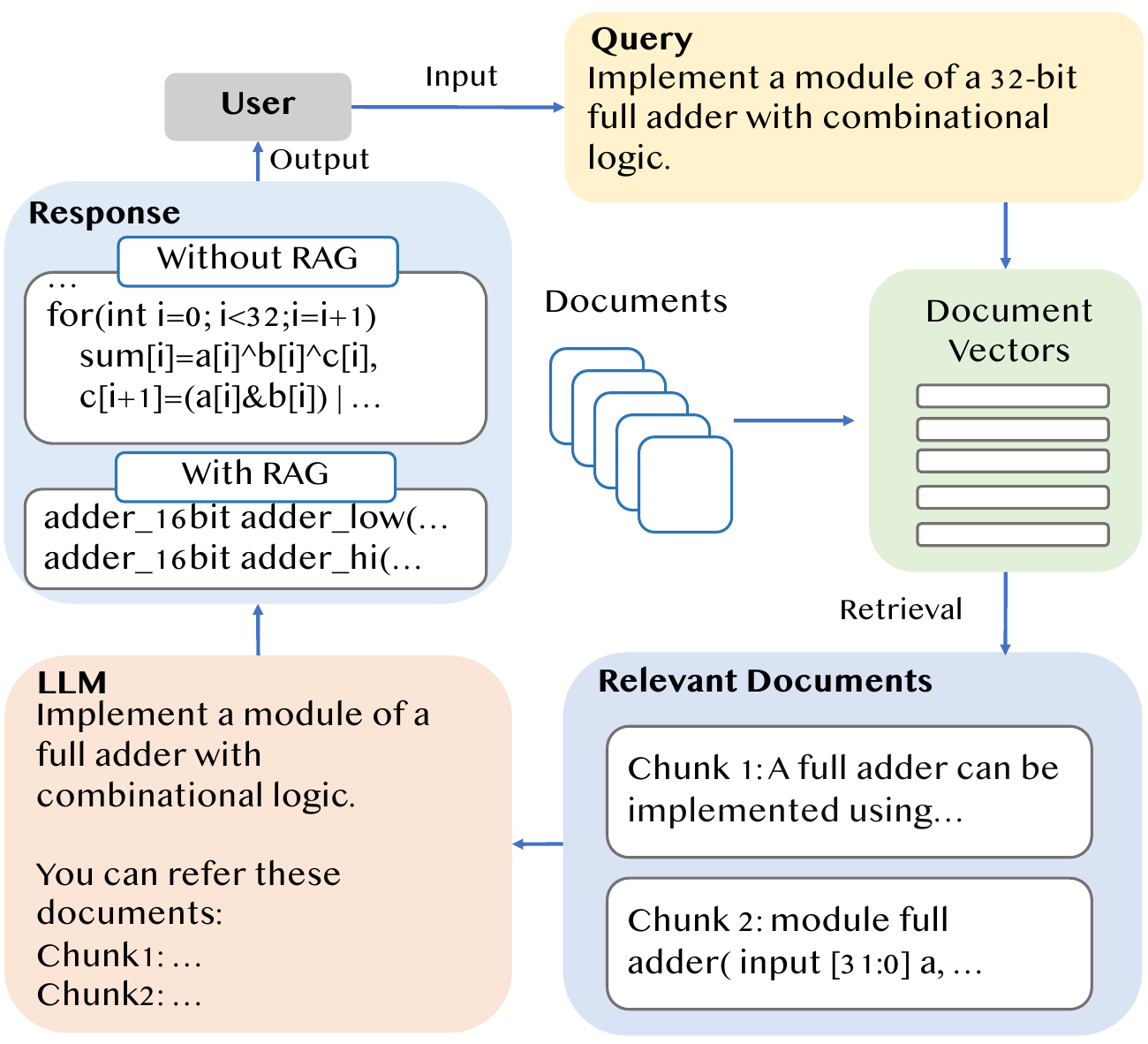} 
	 \caption{An example of the RAG process. }
  \vspace{-0.7cm}
\label{fig_rag_prin} 
\end{figure}

\subsection{Retrieval-Augmented Generation (RAG)}
To address the limitations of LLMs in domain-specific applications, the concept of Retrieval-Augmented Generation (RAG) is introduced, as shown in Fig. \ref{fig_rag_prin}. The user starts by submitting a query, then the RAG module compares this query to chunks in the document database to find the similar chunks. After selecting the chunks, it combines them with the user's query and sends this combined query to LLM for inference. RAG is particularly effective in solving domain-specific problems, and the documents may provide solutions and can ease the issue of hallucinations in LLMs. 
%\textcolor{red}{not convincing.} This approach is particularly effective in contexts where a deep domain knowledge is required, such as hardware code generation.
The RAG module enhances a standard LLM by integrating a retriever that queries a database for domain-specific documents or code snippets during the generation process. This retriever, which is called sentence embedding, usually act as a BERT-like model and can convert sentences into vectors. It is trained to fetch relevant information based on the input prompt.
%using the sentence embedding models we described earlier. 
The retrieved information is then fed back to LLM, providing additional context to generate more appropriate outputs.

In our framework, when generating Verilog code, the RAG module can access examples of similar hardware modules or specific implementations, aiding the LLM in understanding the unique requirements and constraints of hardware design. This also helps LLM to adopt commonly used design patterns, thus enhancing the quality and usability of the generated code. By integrating the retrieval process into the generation pipeline, the RAG model effectively narrows the gap between general-purpose language understanding and the specialized knowledge required for tasks like Verilog code generation. This makes RAG a powerful tool for extending the capabilities of LLMs beyond traditional text generation into more specialized and technical domains.

\section{Methodology}

\begin{figure*}
        \centering
	 \includegraphics[width=0.9\linewidth]{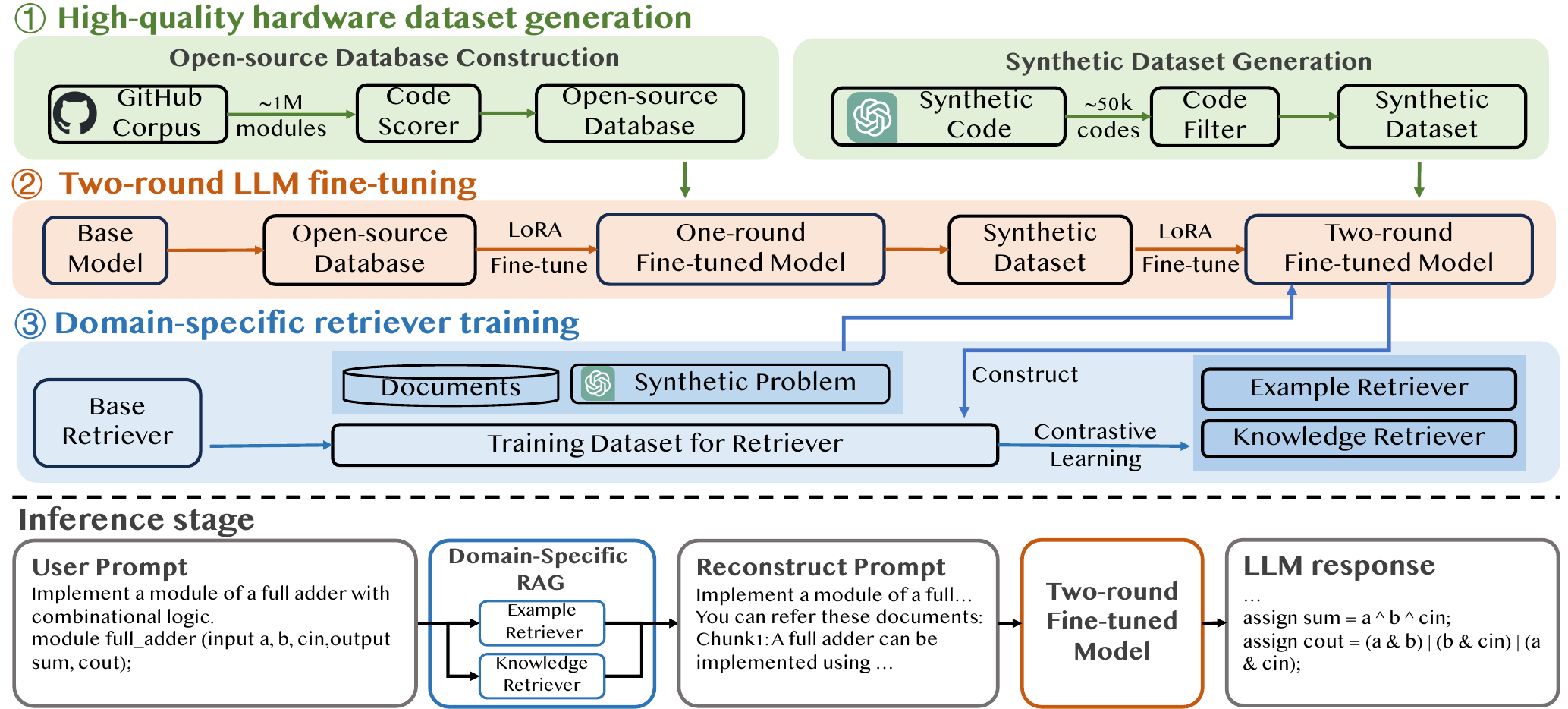} 
	 \caption{Framework overview of AutoVCoder.}
  \vspace{-0.5cm}
\label{fig_overview} \end{figure*}

% Fig. \ref{fig_overview} presents an overview of our framework, which can be divided into three core parts: 1) Training phase of the LLM, 2) Training phase of the RAG retriever, and 3) The inference phase where users interact with the LLM.

Figure \ref{fig_overview} depicts the overview of our framework, which can be divided into three core parts: \ding{182} a high-quality hardware dataset generation approach; \ding{183} a two-round fine-tuning method for LLMs; and \ding{184} a domain-specific retriever training mechanism for RAG.

Firstly, to generate a high-quality hardware dataset in an efficient way, we collect a large number of raw Verilog-based hardware designs from GitHub, after which a novel scoring mechanism (code scorer in Fig. \ref{fig_overview}) that swiftly rates the quality of each hardware design is presented. 
After cleansing data by removing code with low scores, the generated dataset will be used in the first round of LLM fine-tuning to augment LLMs with the capability of understanding the basic syntax of Verilog and the rationale of hardware designs. To further enhance the performance of LLMs on the specific task of generating a correct hardware design given a problem description, we leverage ChatGPT-3.5 to obtain a series of problem-code pairs and propose a verification mechanism (code filter in Fig. \ref{fig_overview}) to ensure their correctness. These samples after filtering form our synthetic dataset which will be used in the second round of LLM fine-tuning.

%During the training of the language model, we engage in two rounds of fine-tuning. The first round involves fine-tuning on code from GitHub, utilizing a novel scoring mechanism that swiftly rates and filters GitHub code. In the second round, we fine-tune on data generated by ChatGPT-3.5. Here, we implement a verification mechanism to ensure the generated data is predominantly correct. This round utilizes problem-code pairs to align with our practical needs. Both rounds of fine-tuning leverage the LoRA\cite{hu2022lora} framework, which helps reduce the time required for fine-tuning and makes the process feasible.

After constructing datasets, we present a two-round LLM fine-tuning method to improve the LLM's efficacy for generating Verilog designs. Starting with a general LLM as the base model, 
we perform the first round of fine-tuning on our generated dataset from open-source GitHub repositories and perform the second round of fine-tuning on the synthetic dataset obtained from ChatGPT-3.5.
%the first fine-tuning stage instruments the model with the large amount of datasets obtained from the GitHub. Subsequently, the second fine-tuning stage leverages the synthetic datasets derived from ChatGPT-3.5, and utilizes problem-code pairs to align the LLM with our practical needs during Verilog generation. 

%During the training of the RAG retriever, we collect Verilog-related materials from textbooks and websites, segmenting them into chunks. We classify these chunks as either knowledges or samples based on the presence of code, thus forming corresponding databases. We also gather domain-specific knowledge, such as RISC-V, from specialized websites to construct dedicated databases. We propose a novel interaction scheme with the trained LLM to form our retriever training dataset, which is then trained using contrastive learning techniques.

%\textcolor{red}{Finally,}

Finally, we utilize the advanced RAG technique to further enhance syntactic and functional correctness during Verilog code generation. We propose a domain-specific retriever training mechanism based on contrastive learning and construct two types of retriever, namely example retriever and knowledge retriever, to fetch different kinds of information. 
% we propose a contrastive learning technique for effective retriever training.
% domain-specific prompt retriever using the advanced RAG technique to guarantee high syntactic and functional correctness during Verilog generation. 
% It is worth noting that we construct two types of retrievers, i.e., example retrieve and knowledge retrieve, and moreover, 
% we propose a contrastive learning technique for effective retriever training.

%During the inference phase, users should accurately describe their problems and the corresponding interface circuits. The system searches through the RAG's document library to find suitable knowledges and samples. The user's prompt is then combined with the retrieved documents to form a reconstructed prompt, which is processed by our LLM to generate the final response.

%\textcolor{red}{check the comments.}
During inference, users can directly describe their problems and ask our model to generate desired Verilog code. As shown in Fig. \ref{fig_overview}, the user prompt is first sent to our domain-specific RAG module which searches through document database and find highly relevant examples and RTL design principles. Then the RAG module reconstructs the input prompt with retrieved contents and feeds this new prompt to the two-round fine-tuned LLM. A high-quality RTL design can then be generated.

% During inference, to deploy our developed LLM, users should produce their problems and the corresponding interface circuits using natural language description. Correspondingly, AutoVCoder searches through the document library of RAG to find highly relevant knowledge and samples as retrieved documents. These retrieved outcomes are combined with the user-provided descriptions to form the prompts, which is later fed into the fine-tuned LLM to generate the final responses to the user problems.

\subsection{Hardware Dataset Generation}
% \paragraph{Collection and filtering of datasets} We utilize a python library GitHub to gather data, which includes around 20,000 repositories. We download these repositories, identified the .v files, and segment them into different module blocks to serve as samples. To ensure the effectiveness of the training data, it is crucial to filter the code. Manual selection is impractical due to the lack of a standard for high-quality code and the extensive time required to sift through such a large dataset. Therefore, we plan to employ LLM to filter the data, we emphasize that the training data should have educational value. Here we choose ChatGPT-3.5 as our "teacher", we add prompts to each module block as shown in fig. \ref{fig_1_prompt}, where the \verb|$code snippet| refers to the code of the module.

\textbf{Open-source database construction.} Hardware designs described in Verilog are usually regarded as valuable assets to each company as well as individual. Therefore, high-quality Verilog-based hardware designs are scarce resources, which makes data-driven learning methods, e.g., LLM, more difficult in the Verilog coding domain. To tackle this problem, we seek to construct a high-quality Verilog design database from the public. We search for open-source RTL code from GitHub, identify .v files in repositories, and segment them into separate blocks to form realistic training samples. Specifically, we gather data from up to 20,000 GitHub repositories and obtain around 1,000,000 raw RTL hardware modules. 

\begin{figure}
        \centering
	 \includegraphics[width=\linewidth]{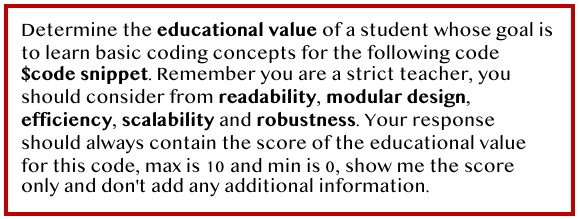} 
	 \caption{Prompt for marking input code with a score. }
   \vspace{-0.3cm}
\label{fig_1_prompt} \end{figure}

Noticing that the online resource is a mixed bag, it is crucial to filter out inferior design cases to maintain the training data's quality. However, due to the large size of the online database, it is impractical to manually look into each design instance and assess its suitability for model training. One way is to utilize ChatGPT-3.5 to search for useful training data instead of going through the process manually. To equip ChatGPT-3.5 with the ability of data cleaning, we add prompts to force ChatGPT-3.5 to behave as a code scorer, with special considerations for Verilog readability, scalability, the degree of standardization, efficiency and robustness, as shown in Fig.~\ref{fig_1_prompt}, where the \verb|$code snippet| refers to the input code.

%With this approach, we utilize ChatGPT-3.5 to obtain scoring information for approximately 15,000 modules. However, scoring all modules, which number over a million, using ChatGPT-3.5 alone is prohibitively expensive and time-consuming. To address this, we design a data filter. Fig. \ref{fig_githubfilter} demonstrate the training and inference step of the scorer, it contains a sentence-transformer module, which is known as a BERT-like neural network , and an MLP layer. We employ the tool known as FlagEmbedding for sentence embedding. In this data filter module, the parameters of the sentence embedding module are fixed, and only the MLP layer is trained using the dataset of 15,000 modules’ code and their corresponding scores obtained via ChatGPT-3.5. Once training is completed, this data filter module enable us to score the remaining modules, providing a preliminary rating for all the code on GitHub. Finally, we assess the quality of the scoring by manually inspecting and evaluating the percentage of data. We found that modules scoring above 6.5 are of higher educational value, comprising about 21.7\% of the total samples. These selected modules are then compiled into a dataset for model training, approximately 340MB in size.

\begin{figure}
        \centering
	 \includegraphics[width=\linewidth]{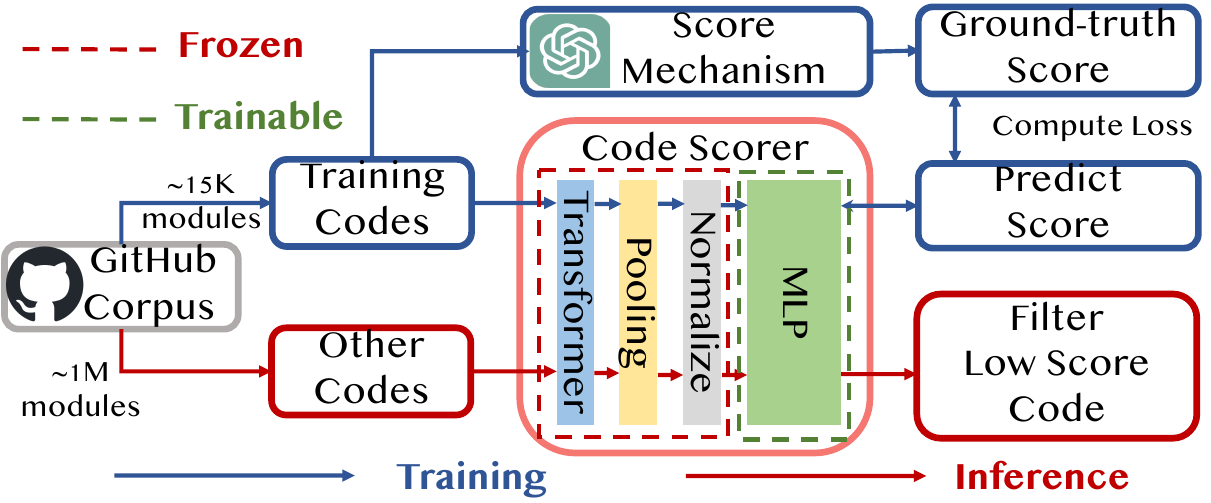} 
	 \caption{Code scoring mechanism with ChatGPT-3.5.}
  \vspace{-0.6cm}
\label{fig_githubfilter} \end{figure}

To speed up the code scoring process and reduce the cost of ChatGPT-3.5, we implement a specialized code scorer as a light-weight replacement of ChatGPT-3.5. Figure \ref{fig_githubfilter} illustrates the training and inference process of our code scorer. It contains a sentence-transformer module and an MLP layer. We employ FlagEmbedding~\cite{llm_embedder} for sentence embedding. In this code scoring model, the parameters of the sentence-transformer module are fixed, and only the MLP layer is trainable. During the training stage, we update the MLP layer with only a subset of the complete open-source database, i.e., 15,000 Verilog modules and their corresponding scores obtained via ChatGPT-3.5. Once training is completed, the code scorer is applied to score the remaining Verilog modules within the open-source database. We found that Verilog modules scoring above 6.5 are of high educational value, accounting for about 21.7\% of the total samples. Hence, these high-score Verilog modules are used for the first-round LLM fine-tuning. 

% \paragraph{Collection and filtering of datasets}

% During the data collection phase, we aim to generate high-quality problem-code pairs to standardize our LLM for problem solving. We continued to use ChatGPT-3.5 to help us create high-quality synthetic data. Initially, we let ChatGPT-3.5 generate a problem related to Verilog, followed by having it provide the answer. To ensure the generated code was varied, we employ a simple strategy known as tinystory\cite{eldan2023tinystories}. This method involves altering a few words in the prompt to change its structure, thereby ensuring diversity in the code. The format for this prompt is shown in fig. \ref{fig_2_prompt}, where, \verb|$level| refers to easy, normal or hard, \verb|$circuit type| refers to either combinational or sequential logic circuits, and \verb|$problem type| denotes a type of problem that will be randomly selected from a list of problem types.

\begin{figure}
        \centering
	 \includegraphics[width=\linewidth]{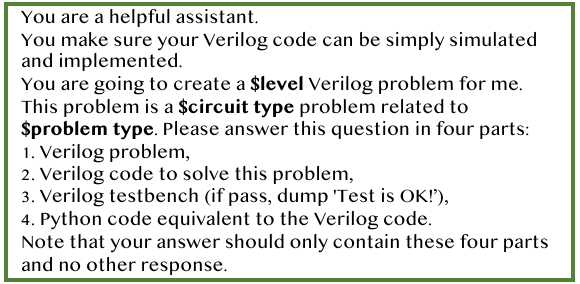} 
	 \caption{Prompt for generating problem-code pairs. }
   \vspace{-0.3cm}
\label{fig_2_prompt} \end{figure}

\textbf{Synthetic dataset generation.} Besides using open-source database to learn basic RTL syntax, we also seek to generate practical problem-code pairs to standardize our LLM in the problem solving task. We continued to use ChatGPT-3.5 to create a series of specialized and synthetic data. Initially, we let ChatGPT-3.5 generate a problem related to Verilog, and at the same time, we ask ChatGPT-3.5 to provide the answer. To ensure the diversity of the generated code, we learn from the tinystory strategy~\cite{eldan2023tinystories} and increase the code variety by changing several keywords in the prompt, as shown in Fig.~\ref{fig_2_prompt}, where \verb|$level| refers to easy, normal or hard, \verb|$circuit type| refers to either combinational or sequential logic circuits, and \verb|$problem type| denotes a type of problem that will be randomly selected from a list of problem types.

\begin{figure}
        \centering
	 \includegraphics[width=\linewidth]{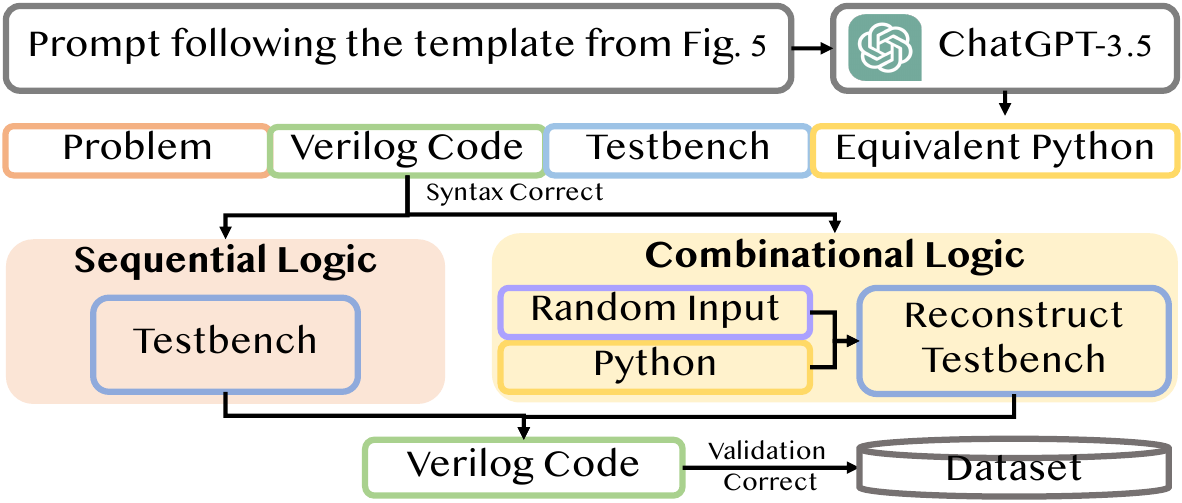} 
	 \caption{The flow of our code filter.}
  \vspace{-0.5cm}
\label{fig_chatgptfilter} \end{figure}

However, it's important to note that the Verilog code generated by ChatGPT-3.5 may not always be correct. That is, it can incur syntactic or functional errors. To address this issue, we design a code filter to help identify invalid code samples. As shown in Fig. \ref{fig_chatgptfilter}, we use ChatGPT-3.5 to generate four components: the problem, the Verilog code, the corresponding testbench, and the equivalent Python code. We first utilize the Icarus Verilog tool~\cite{aslan2016open} to verify the syntax correctness of the generated code. If the code is syntactically correct, we further evaluate its functional correctness. We note that Verilog code generally falls into two categories: combinational and sequential logic circuits. For combinational circuits, we utilize the equivalent Python code to conduct auxiliary checks. We generate random inputs for the Python code, obtain corresponding outputs, and use these outputs to reconstruct and create testbenches for the Verilog code, which we then synthesize and analyze. For sequential circuits, we use the testbench created by ChatGPT-3.5 to check if expected outputs are provided. Different approaches are taken because LLM has a strong capability in analyzing sequential information, making it easier to understand and write testbenches for sequential logic circuits. However, its capacity for numerical calculation is limited, leading to errors when writing testbenches for combinational logic circuits. Therefore, we need to reconstruct the testbench for combinational logic to ensure the correctness.
% However, a limitation of this method is that errors may exist when writing testbenches for combinational logic circuits. This problem can be alleviated by forcing ChatGPT-3.5 to generate more synthetic datasets.

%\subsection{Fine-tune LLM with GitHub Code}
\subsection{Two-Round LLM Fine-Tuning}

%We start with an existing open-source LLM as our base model and fine-tune it. The initial round of fine-tuning is designed to help the LLM learn the syntactic structure of Verilog code and its practical uses in production environments. The objective is to refine the model so it can proficiently generate Verilog code. During this process, we emphasize two key steps: the collection and filtering of datasets, and the training process.

%\paragraph{Training Process}

%In the initial phase of our model training, we choose to use the LoRA method to fine-tune LLM. This approach provides quicker results compared to traditional full-parameter fine-tuning. During this process, we apply the Maximum Likelihood Estimation (MLE) loss function alongside LoRA fine-tuning methods. We specifically focus on analyzing the probability differences between the generated text and the actual text to improve accuracy, where we use the cross-entropy loss to formula to measure discrepancies more precisely.

\textbf{Fine-tuning with the open-source database.} We start with a well-trained LLM as our base model and fine-tune it with the open-source database. The first fine-tuning stage is designed to help the LLM learn the syntactic structure of Verilog code and its practical uses in production environments. In this initial phase of model fine-tuning, we adopt the low rank adaptation(LoRA)~\cite{hu2022lora} method. This approach provides results faster compared to traditional full-parameter fine-tuning. We apply the Maximum Likelihood Estimation (MLE) loss function alongside the LoRA~\cite{hu2022lora} fine-tuning method. Regarding the loss function, we use the cross-entropy loss to quantify the discrepancies between the generated text and the actual text.

%\subsection{Fine-tune LLM with Synthetic Data}

%After the first round of fine-tuning, we developed an LLM capable of generating reasonable code. To enhance this further, in the second round of fine-tuning, we aim to refine our methods to improve the LLM's performance in practical production settings. During this phase, we introduce an innovative data collection strategy designed to gather high-quality datasets, ensuring that our model becomes even more effective and reliable. As the same as the previous process, we also emphasize the collection and filtering of datasets, and the training process in this process.

%\paragraph{Training Process}

%The second round of fine-tuning used an instruction tuning approach, where we treat the initial problems as instructions and the code outputs as tasks. We utilized ChatGPT-3.5 to generate over 10,000 high-quality problem-code pairs. Building on the model that was fine-tuned in the first round, we continue training using the same structure as the first round, employing the LoRA method to adjust the parameters of LLM.

\textbf{Fine-tuning with the synthetic dataset.} In the second round of fine-tuning, we aim to improve the LLM's performance in more realistic scenarios. We fine tune the model to perform the specific QA task: providing an answer with correct Verilog code to a hardware design problem. Specifically, we use an instruction tuning approach to standardize the outputs of our model. This ensures that when our model receives Verilog problem, it can produce the code that corresponds to that problem.
%Specifically, we use an instruction tuning approach, where we treat the initial problems as instructions and the code outputs as tasks 
% \textcolor{red}{what do you mean "treat code output as tasks"}. In this stage, we make use of over 10,000 high-quality problem-code pairs in the generated synthetic dataset.

\begin{figure*}
        \centering
	 \includegraphics[width=0.97\linewidth]{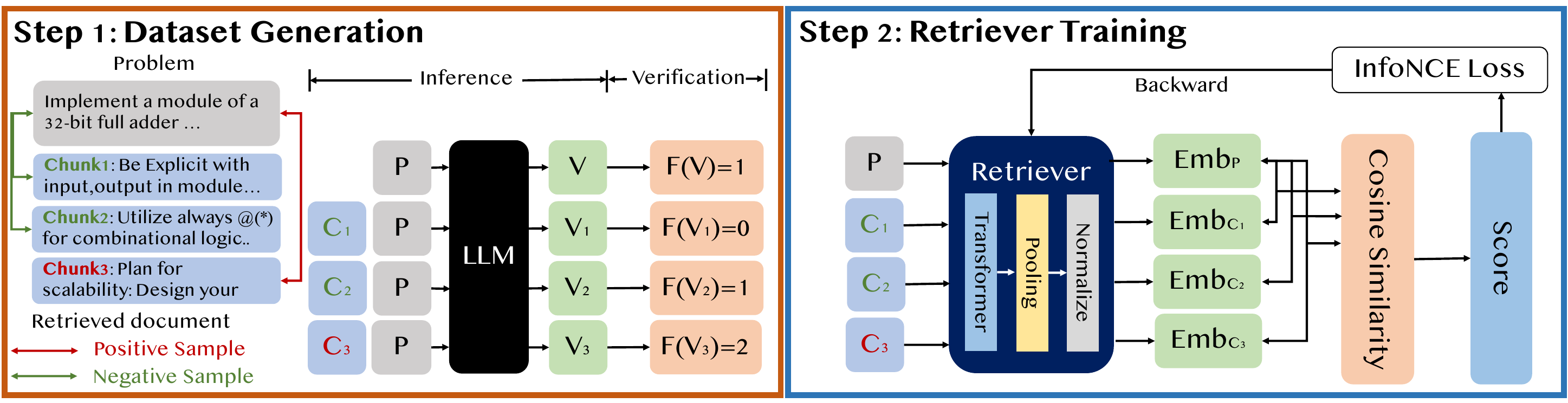} 
	 \caption{The process of constructing our domain-specific retrievers for RAG.}
  \vspace{-0.5cm}
\label{fig_rag} \end{figure*}

\subsection{Domain-Specific Retrieval-Augmented Generation}

%RAG enables the model to extract relevant information from external documents dynamically, improving the ability of LLM to generate code that is more accurate and contextually appropriate. To fetch useful information efficiently, we propose a domain-specific RAG process, consisting of two retrievers: the example retriever and the knowledge retriever. We utilize these two retrievers to search different kinds of information. 
% Specifically, these retrievers seek specialized information as outlined below. 

The rationale of RAG is to identify the piece of data in an existing database that is most correlated to the current task of interests. The extracted relevant data is then used as additional prompt to improve the ability of LLM to generate code that is more accurate and contextually appropriate. In order to search for the useful information in an efficient manner, we propose a domain-specific RAG scheme, which consists of two types of retrievers, i.e., an example retriever and a knowledge retriever that fetch different types of domain-specific information correspondingly.

%The knowledge retriever is primarily concerned with understanding what actions LLM should take and should not take when performing certain tasks. For instance, a significant flaw in traditional LLMs may have a tendency to favor the use of loops like Python, C++, and Java. This tendency often leads to excessive usage of for loops when generating code, which can be resource-intensive and not reflective of actual coding practices. Therefore, providing prompts that discourage the indiscriminate use of for loops is crucial. In many cases, by offering subtle hints, we can help the LLM recognize and correct its unreasonable errors, thereby improving its output. In practical terms, we extract paragraphs from a large corpus of Verilog textbooks and blogs to obtain knowledge-related information about Verilog. Then, using this knowledge retriever, we identify several relevant pieces of knowledge to incorporate into the prompt for the LLM, ensuring its ability to generate code more effectively.

\textbf{Example retriever.}
%The example retriever searches documents and provides pre-cached demonstration examples. It can identify several examples that closely match the problem description. Then the example retriever combines these examples with the user's question and inputs them into the LLM for inference. The purpose of this approach is to enable LLM to perform in-context learning with these examples. By learning from several cases, the model can enhance its understanding of the user's question and generate hardware code with higher quality. For instance, when designing the circuit for a traffic light, we often use a Finite State Machine (FSM). The example retriever would provide several examples of FSM for the LLM to learn. In practice, our examples are extracted from various reliable sources such as textbooks, blogs, and specialized websites dedicated to FPGA and ASIC design. These diverse sources ensure that the model is exposed to a wide range of contexts and applications within the field.
The purpose of applying the example retriever is to enable the LLM to perform in-context learning with these given examples. Specifically, the example retriever searches from a document database, and pinpoints the demonstration examples that closely match with the problem description. Then, these examples, together with the user's question, are fed into the LLM for inference. By learning common knowledge from these highly relevant cases, the model is augmented to better understand the user's intention and tends to generate hardware code with higher quality. For instance, the finite state machine (FSM) is a widely used computation model for describing states and their transitions in sequential logic design. When the input question opts for a FSM design, e.g., a traffic light, the example retriever would provide several examples of FSM for the LLM to learn more efficiently. In practice, our examples are extracted from a database constructed with various reliable sources such as textbooks, blogs, and specialized websites dedicated to FPGA and ASIC design. The diversity of database sources ensures that the model can have access to a wide range of contexts and applications within the required field.

\textbf{Knowledge retriever.} 
%The knowledge retriever provides external knowledge to support LLMs in tackling knowledge-intensive tasks and ease the issue of hallucinations. It can find paragraphs or principle information that closely relate to the problem description. For example, a significant flaw in traditional LLMs is their tendency to use loops like those in Python and C++, as shown in the example in Fig. \ref{fig_rag_prin}. This tendency often leads to excessive usage of \texttt{for} loops when generating RTL code, which can be resource-consuming and does not conform to realistic RTL coding practices. Therefore, providing knowledge that discourage the excessive use of \texttt{for} loops is crucial to solve this software hallucination problem.
The knowledge retriever extracts RTL design principles and supplementary descriptions about key terminologies in the question to support LLMs in tackling knowledge-intensive tasks. It aims to find paragraphs or principles closely related to the problem description. Our knowledge retriever can help ease the issue that the generated output is grammatically correct but does not obey the RTL design rationale, which is known as \emph{hallucinations of LLM}. For instance, we observe that a significant flaw in traditional LLMs is their tendency to use loops like those in Python and C++, as shown in the example in Fig. \ref{fig_rag_prin}. This tendency often leads to excessive usage of \texttt{for} loops when generating RTL code, which can be resource-consuming and does not conform to realistic RTL coding practices. Therefore, providing knowledge that discourages the excessive use of \texttt{for} loops is crucial to solve this software hallucination problem.

On the other hand, gathering constructive examples is challenging because it requires finding formally structured code and verifying its correctness before it can be included in the example document, while collecting knowledge chunks is relatively easier. Therefore, the knowledge retriever can serve as a supplement to our example retriever. In practice, we build our knowledge database by extracting paragraphs from a large corpus of Verilog textbooks and blogs to obtain information about Verilog and hardware terminologies, given the assumption that the knowledge collected from textbooks and domain-specific websites is correct. For example, when the LLM is asked to design a Booth multiplier using Verilog, it is highly possible that the example retriever cannot effectively find valid examples due to the fact that the Booth multiplier is a specific terminology dedicated to hardware design and the LLM may not have knowledge about it beforehand. In this situation, the knowledge retriever can extract from the textbooks the definition and functionality of the Booth multiplier, which is fed into the LLM for learning. 

%The sample retriever enables LLM to better learn the underlying representations of examples through few-shot learning. By employing in-context learning, we can provide examples simply through few-shot methods. Using examples is essential as they provide concrete instances for the model to learn from, aiding in better understanding the underlying representations of concepts. By incorporating in-context learning, we simplify the provision of examples, making it easier to impart knowledge to the model using only a few instances. Our examples are sourced from various reliable sources such as textbooks, blogs, and specialized websites dedicated to FPGA and ASIC design. These diverse sources ensure that the model is exposed to a wide range of contexts and applications within the field, enriching its understanding and improving its ability to generate accurate and relevant code.

%\paragraph{Domain Expertise Retriever}

%The domain expertise retriever is closely linked to RISC-V. Its role is to gather specialized knowledge concerning the RISC-V architecture, including details about instruction sets, pipeline stages, and memory organization. This information is vital for guiding the model's comprehension and generation of RISC-V code. Similarly, we've developed a dedicated document containing structural information and Q\&A for RISC-V. The aim is to validate our large language model's ability to extend to practical applications. 

\textbf{Construction of retrievers.}
A key challenge is how to construct retrievers effectively to ensure the retrieved information is relevant and accurate. A simple method is to represent problems and document contents with general sentence embeddings and compare their differences for similarity evaluation. However, the performance would be influenced 
% Simply using general sentence embeddings and then comparing differences between sentence embeddings for similarity evaluation are not sufficient, 
because it is hard for someone with limited RTL background to find relevant information just from a problem description. 
Therefore, additional training of retrievers is required to ensure questions match up well with the retrieved information. 
The establishment of retrievers is divided into two steps: \ding{182} dataset generation and \ding{183} retriever training, as illustrated in Fig. \ref{fig_rag}.

\textit{Step 1: dataset generation.} %We use the two-round fine-tuned Codellama (referred to as $\mathbb{M}$ after) for dataset preparation of the retriever training. We additionally generate 2,000 correct Verilog code problems using ChatGPT-3.5, similar to our synthetic dataset generation. The model and problem sets help us build our retriever dataset.
%The generation of the dataset relies on two key components: an LLM $\mathbb{M}$ capable of generating Verilog code and a Verilog problem set $\mathcal{P}$. For the LLM $\mathbb{M}$, we followed the steps in the methodology, using the Codellama-7B\cite{rozière2024code} as base model and fine-tune it for two rounds. For Verilog problem set $\mathcal{P}$, we generate additional 2,000 correct Verilog code problems using ChatGPT-3.5, which are similar to those in our synthetic dataset.
To train our retrievers, we utilize contrastive learning which requires a dataset with a large number of positive and negative sample pairs. If the addition of a document chunk enhances the output quality of an LLM for a given question, this document chunk and the question form a positive sample pair. Conversely, if the addition of a document chunk fails to improve or even reduces output quality, this document chunk and the question form a negative sample pair. Following this criterion, we can combine a question $P$ with our documents to generate multiple sets of positive and negative sample pairs.
% Taking the dataset generation process for training the example retriever as an example, for simplicity, 
To automate the dataset generation process, we first define a function $F$ to evaluate the degree of accuracy:
\[
F(V) = 
\begin{cases} 
0, & \text{syntax}\text{ is incorrect} \\
1, & \text{syntax}\text{ is correct but functionality is incorrect}\\
2, & \text{syntax and functionality}\text{ are correct} 
\end{cases} 
\]
where $V$ represents a Verilog code snippet. 

% Figure \ref{fig_rag} illustrates the process of generating positive and negative samples for our dataset.
 %We define a problem \( P \), an LLM model $\mathbb{M}$ for Verilog code generation, and a judgment function \( F \) for Verilog code. Given a Verilog code \(V\), we define function \( F \):
As shown in Fig. \ref{fig_rag}, given a problem $P$, an LLM is first utilized to generate a Verilog code solution $V$, which works as the comparison baseline to differentiate positive and negative samples. To be more specific, after retrieving multiple document chunks, each chunk is combined with the problem to form a new prompt which is processed by LLM to generate the Verilog code solution, $V_i$, correspondingly. All the generated Verilog code outputs are tested with the Icarus Verilog tool~\cite{aslan2016open} to evaluate their accuracy, i.e., $F(V_i)$. Then the positive and negative sample pairs can be classified by comparing $F(V_i)$ and $F(V)$.
If $F(V_i)>F(V)$, it indicates that adding this chunk benefits the RTL code generation and hence we record this chunk and problem as a positive sample pair. If $F(V_i)<=F(V)$, it suggests that adding this chunk has a negative or no impact on the RTL code generation, and we record them as a negative sample pair. 
% If $F(V_i)=F(V)$, it means this chunk has no significant impact on this problem. In this case, this chunk and problem is also considered as a negative sample pair. 
% It is worth nothing that when \( F(V) = F(V') = 2 \), we cannot judge whether the chunk enhances problem or not. Therefore, we discard the sample pairs when \( F(V) = F(V') = 2 \).
Following this strategy, a large set of positive and negative samples (around 200,000 samples in total) can be generated automatically, which will be used for retriever training.

% We use the method described above to generate a large number of positive and negative samples, which we then use to train our retriever. 

 %we first utilize $\mathbb{M}$ to obtain the corresponding output $V = \mathbb{M}(P)$. We record the corresponding \( F(V) \). Then, we find all the example in the example document. Take the i-th example $E_i$ as an example, we transform this problem into a prompt \( Pr = \text{concat} (E_i, P) \). We validate the result  $V' = \mathbb{M}(Pr)$. Clearly, if \( F(V') > F(V) \), it indicates that adding the example  \( E_i \) has significantly helped with the generation of this problem. We record \( ( E_i,P) \) as a positive sample pair. If \( F(V') < F(V) \), it suggests that adding the example has a negative impact on generating this problem, and we record \( (E_i,P) \) as a negative sample pair. If \( F(V') = F(V) \), it means \( E_i \) has no significant impact on problem \( P \). In this case, \( (E_i,P) \) is also considered a negative sample pair, as we do not want to encounter \( E_i \) when asking problem \( P \). It is worth nothing that when \( F(V) = F(V') = 2 \), we cannot judge whether \( E_i \) enhances problem \( P \). This \( E_i \) may be meaningful example or irrelevant example. Therefore, we discard text sample pairs when \( F(V) = F(V') = 2 \). We use the method described above to generate a large number of positive and negative samples, which we then use to train our retriever. 

\textit{Step 2: retriever training.} We adopt FlagEmbedding~\cite{llm_embedder} as our base retriever and fine-tune it on our dataset for our task. We employ contrastive learning and the training process is illustrated in Fig. \ref{fig_rag}. The InfoNCE loss function~\cite{oord2018representation} is adopted, which is formulated as follows:
$$
 Loss = \sum_{(P,C_+)} -\log \frac{\exp ( \langle e_P, e_{C_+} \rangle / \tau)}{\sum_{C_- \in \mathcal{D}} (\exp \langle e_P, e_{C_-} \rangle / \tau)},
$$
where $P$ denotes a problem, $C_+$ denotes a positive chunk, $C_-$ denotes a negative chunk, $e_i$ denotes the sentence embedding of a sentence $i$, $\tau$ represents temperature, and $\mathcal{D}$ refers to documents. $\langle \rangle$ denotes the dot product of vectors. Since both vectors are normalized, their dot product represents the cosine value between them. 
To minimize the loss during contrastive learning, we aim to maximize the cosine value of positive sample pairs (the numerator), while minimizing the cosine value of negative sample pairs (the denominator). 

Both our example retriever and knowledge retriever are trained following the process in Fig. \ref{fig_rag}.
% We employ contrastive learning with $Loss$ to train our example retriever and knowledge retriever. 
After constructing the retrievers, we can deploy them to fetch relevant and useful information from documents. During inference, the number of retrieved chunks for each problem can be determined by users. In our experiments, we set the numbers of chunks retrieved by example retriever and knowledge retriever to two and three, correspondingly, considering both efficiency and accuracy. The retrieved chunks along with the problem form the prompt to help LLM achieve in-context learning and generate high-quality RTL code.
% our retrievers 
% For inference stage, we find 3 $\sim$ 5 useful knowledge chunks, along with 1 $\sim$ 2 examples, to form our prompt, enabling our model to engage in-context learning, and then output a result for the corresponding question. 
Our experiments show that the RAG module effectively improves the quality of generated Verilog code.

%For domain expertise retriever, the problems we want to understand are often related to the documents themselves. Therefore, we do not need to retrain, but only need to use general sentence embedding for retrieval. In actual application, we first determine whether the question asked belongs to domain expertise. This can be judged by analyzing the similarity between the documents in domain knowledge and the question. If it does, we intervene in the domain knowledge retriever; if not, we skip this step. 

\vspace{-1mm}
\section{Experiments}
%Our experiment focuses on several aspects of our framework. Firstly, we aim to understand the capabilities of the baseline model and explain the necessity and benefits of conducting two rounds of fine-tuning. We also discuss the advantages this framework brings to our project. Second, we need to analyze the advantages of our two-stage fine-tuning model compared to SOTA RTL code generation models. Finally, we need to evaluate the improvements our introduced RAG module contributes to our tasks after fine-tuning.

We conduct a series of experiments to showcase the advancement of the proposed framework, AutoVCoder. Firstly, we perform the end-to-end comparison, evaluating the syntactic and functional correctness achieved by our framework, and comparing the results with state-of-the-art (SOTA) methods from both the industry and academia.
Secondly, we evaluate the improvement of our two-round fine-tuning strategy and we perform ablation studies to examine the efficacy of each round of fine-tuning over the base LLMs and recent models. Thirdly, the proposed domain-specific RAG techniques, i.e., the example retriever and the knowledge retriever, are tested under various experimental settings.

\begin{table*}[ht]
\centering
\caption{Comparison between AutoVCoder and the state-of-the art methods.}
\vspace{-1mm}
\begin{tabular}{|c|>{\centering\arraybackslash}p{1cm}|>{\centering\arraybackslash}p{1cm}|>{\centering\arraybackslash}p{1cm}|>{\centering\arraybackslash}p{1cm}|>{\centering\arraybackslash}p{1cm}|>{\centering\arraybackslash}p{2.5cm}|>{\centering\arraybackslash}p{2.5cm}|}
\hline
\multirow{3}{*}{Evaluated Model} & \multicolumn{1}{c|}{\multirow{3}{*}{\begin{tabular}[c]{@{}c@{}}Num \\ of Params\end{tabular}}} & \multicolumn{4}{c|}{\begin{tabular}[c]{@{}c@{}}VerilogEval Benchmark (using pass@k metric)\end{tabular}} & \multicolumn{2}{c|}{\begin{tabular}[c]{@{}c@{}}RTLLM V1.1 (using pass@5 metric)\end{tabular}} \\ \cline{3-8} 
                                 & \multicolumn{1}{c|}{}                                                                          & \multicolumn{2}{c|}{EvalMachine}                     & \multicolumn{2}{c|}{EvalHuman}                      & \multicolumn{1}{c|}{\multirow{2}{*}{Syn.}} & \multicolumn{1}{c|}{\multirow{2}{*}{Func.}} \\ \cline{3-6}
                                 & \multicolumn{1}{c|}{}                                                                          & \multicolumn{1}{c|}{k=1}  & \multicolumn{1}{c|}{k=5} & \multicolumn{1}{c|}{k=1} & \multicolumn{1}{c|}{k=5} & \multicolumn{1}{c|}{}                          & \multicolumn{1}{c|}{}                           \\ \hline

GPT-3.5 & N/A & 46.7\% & 69.1\% & 26.7\% & 45.8\% &  89.7\% & 37.9\%\\
GPT-4 & N/A & \cellcolor{mygreen!80}\textbf{60.0\%} & \cellcolor{mygreen!80}\textbf{70.6\%} & \cellcolor{mygreen!80}\textbf{43.5\%} & \cellcolor{mygreen!80}\textbf{55.8\%} & \cellcolor{mygreen!80}\textbf{100\%} & \cellcolor{mygreen!80}\textbf{65.5\%} \\

%\midrule
\hline

ChipNeMo~\cite{liu2024chipnemo} & 13B & 43.4\% & N/A & 22.4\% & N/A & N/A & N/A \\
VerilogEval~\cite{liu2023verilogeval} & 16B & 46.2\% & 67.3\% & 28.8\% & 45.9\% & N/A & N/A \\

Codegen2~\cite{nijkamp2023codegen2} & 16B & 5.00\% & 9.00\% & 0.9\% & 4.1\% & 72.4\% & 6.9\% \\
Starcoder~\cite{li2023starcoder} & 15B & 46.8\% & 54.5\% & 18.1\% & 26.1\% & 93.1\% & 27.6\% \\
Thakur et al.\cite{date23} & 16B & 44.0\% & 52.6\% & 30.3\% & 43.9\% & 86.2\% & 24.1\% \\

RTLCoder-Mistral~\cite{liu2024rtlcoder} & 7B & 62.5\% & 72.2\% & 36.7\% & 45.5\% & \cellcolor{yanse!70}\textbf{96.6\%} & 48.3\% \\
RTLCoder-DeepSeek~\cite{liu2024rtlcoder} & 6.7B & 61.2\% & 76.5\% & 41.6\% & 50.1\% & 93.1\% & \cellcolor{yanse!70}\textbf{48.3\%} \\ 

BetterV-Codellama~\cite{betterv} & 7B & 64.2\% & 75.4\% & 40.9\% & 50.0\% & N/A & N/A \\ 
BetterV-DeepSeek~\cite{betterv} & 6.7B & 67.8\% & 79.1\% & 45.9\% & 53.3\% & N/A & N/A \\ 
BetterV-CodeQwen~\cite{betterv} & 7B & 68.1\% & \cellcolor{yanse!70}\textbf{79.4\%} & 46.1\% & \cellcolor{yanse!70}\textbf{53.7\%} & N/A & N/A \\ 

%AutoVCoder-Codellama-1\&2 & 7B & 60.1\% & 66.6\% & 41.6\% & 47.1\% & 89.6\% & 44.8\%\\
AutoVCoder-Codellama & 7B & 63.7\% & 72.9\% & 44.5\% & 52.8\% & 93.1\% & \cellcolor{yanse!70}\textbf{48.3\%} \\

%AutoVCoder-Mistral-1\&2 & 7B & 59.9\% & 69.1\% & 40.5\% & 45.0\% & 96.6\% & 51.7\% \\
AutoVCoder-DeepSeek & 6.7B & \cellcolor{mygreen!80}\textbf{69.0\%} & 79.3\% & \cellcolor{yanse!70}\textbf{46.9\%} & \cellcolor{yanse!70}\textbf{53.7\%} & \cellcolor{mygreen!80}\textbf{100\%}  & \cellcolor{mygreen!80}\textbf{51.7\%} \\

AutoVCoder-CodeQwen & 7B & \cellcolor{yanse!70}\textbf{68.7\%} & \cellcolor{mygreen!80}\textbf{79.9\%} & \cellcolor{mygreen!80}\textbf{48.5\%} & \cellcolor{mygreen!80}\textbf{55.9\%} & \cellcolor{mygreen!80}\textbf{100\%}  & \cellcolor{mygreen!80}\textbf{51.7\%}\\
\hline

\end{tabular}
\label{tab_2}
\end{table*}

\begin{table*}[ht]
\centering
\vspace{-0.1cm}
\caption{Evaluation of various models with different fine-tuning strategies.}
\vspace{-1mm}
\begin{tabular}{|c|>{\centering\arraybackslash}p{1cm}|>{\centering\arraybackslash}p{1cm}|>{\centering\arraybackslash}p{1cm}|>{\centering\arraybackslash}p{1cm}|>{\centering\arraybackslash}p{2.5cm}|>{\centering\arraybackslash}p{2.5cm}|}
\hline
\multirow{3}{*}{Evaluated Model} & \multicolumn{4}{c|}{\begin{tabular}[c]{@{}c@{}}VerilogEval Benchmark (using pass@k metric)\end{tabular}} & \multicolumn{2}{c|}{\begin{tabular}[c]{@{}c@{}}RTLLM V1.1 (using pass@5 metric)\end{tabular}} \\ \cline{2-7} 
                                                                                                        & \multicolumn{2}{c|}{EvalMachine}                     & \multicolumn{2}{c|}{EvalHuman}                      & \multicolumn{1}{c|}{\multirow{2}{*}{Syn.}} & \multicolumn{1}{c|}{\multirow{2}{*}{Func.}} \\ \cline{2-5}
                                 & \multicolumn{1}{c|}{k=1}  & \multicolumn{1}{c|}{k=5} & \multicolumn{1}{c|}{k=1} & \multicolumn{1}{c|}{k=5} & \multicolumn{1}{c|}{}                          & \multicolumn{1}{c|}{}                           \\ \hline

Codellama-7B\cite{rozière2024code} & 34.1\% & 41.3\% & 21.7\% & 24.5\% & 62.1\% & 10.3\%\\
AutoVCoder-Codellama-1 & 39.2\% & 46.7\% & 28.5\% & 31.3\% & 72.4\% & 14.0\%\\
AutoVCoder-Codellama-2 & 55.1\% & 59.9\% & 39.2\% & 43.5\% & 89.6\% & 37.9\%\\
AutoVCoder-Codellama-1\&2 & \textbf{60.1\%} & \textbf{66.6\%} & \textbf{41.6\%} & \textbf{47.1\%} & \textbf{89.6\%} & \textbf{44.8\%}\\

\hline%\midrule

DeepSeek-Coder-6.7B\cite{guo2024deepseekcoderlargelanguagemodel} & 52.1\% & 56.4\% & 30.8\% & 34.2\% & 89.6\% & 34.5\% \\
AutoVCoder-DeepSeek-1 & 57.3\% & 67.7\% & 34.8\% & 38.4\% & 93.1\% & 34.5\% \\
AutoVCoder-DeepSeek-2 & 65.5\% & 74.3\% & \textbf{45.5\%} & 51.8\% & 93.1\% & 44.8\%\\
AutoVCoder-DeepSeek-1\&2 & \textbf{67.1\%} & \textbf{77.8\%} & 45.1\% & \textbf{52.8\% }& \textbf{100\%} & \textbf{51.7\%} \\

\hline%\midrule

CodeQwen-7B\cite{bai2023qwentechnicalreport} & 48.0\% & 52.8\% & 23.2\% & 28.1\% & 82.7\%  & 27.6\%\\ 
AutoVCoder-CodeQwen-1 & 58.9\% & 65.4\% & 32.7\% & 36.1\% & 82.7\% & 34.5\% \\
AutoVCoder-CodeQwen-2 & 65.3\% & 75.1\% & 45.5\% & 51.2\%  & 96.6\%  & 44.8\%\\
AutoVCoder-CodeQwen-1\&2 & \textbf{66.8\%} & \textbf{78.3\%} & \textbf{46.2\%} & \textbf{54.1\%} & \textbf{100\%}  & \textbf{51.7\%} \\\hline
\end{tabular}
\label{tab_1}
\end{table*}

\begin{table*}[ht]
\centering
\vspace{-0.1cm}
\caption{Evaluation of various models with different types of retrievers.}
\vspace{-1mm}
\begin{tabular}{|c|>{\centering\arraybackslash}p{1cm}|>{\centering\arraybackslash}p{1cm}|>{\centering\arraybackslash}p{1cm}|>{\centering\arraybackslash}p{1cm}|>{\centering\arraybackslash}p{2.5cm}|>{\centering\arraybackslash}p{2.5cm}|}
\hline
\multirow{3}{*}{Evaluated Model} & \multicolumn{4}{c|}{\begin{tabular}[c]{@{}c@{}}VerilogEval Benchmark (using pass@k metric)\end{tabular}} & \multicolumn{2}{c|}{\begin{tabular}[c]{@{}c@{}}RTLLM V1.1 (using pass@5 metric)\end{tabular}} \\ \cline{2-7}                                                                          & \multicolumn{2}{c|}{EvalMachine}                     & \multicolumn{2}{c|}{EvalHuman}                      & \multicolumn{1}{c|}{\multirow{2}{*}{Syn.}} & \multicolumn{1}{c|}{\multirow{2}{*}{Func.}} \\ \cline{2-5}                                                                          & \multicolumn{1}{c|}{k=1}  & \multicolumn{1}{c|}{k=5} & \multicolumn{1}{c|}{k=1} & \multicolumn{1}{c|}{k=5} & \multicolumn{1}{c|}{}                          & \multicolumn{1}{c|}{}                           \\ \hline

AutoVCoder-Codellama-1\&2 & 60.1\% & 66.6\% & 41.6\% & 47.1\% & 89.6\% & 44.8\%\\                                 
AutoVCoder-Codellama-ER & 63.1\% & 70.1\% & \textbf{44.5\%} & 51.1\% & \textbf{93.1\%} & \textbf{48.3\%} \\
AutoVCoder-Codellama-KR & 60.9\% & 68.1\% & 40.5\% & 47.9\% & 89.6\% & 44.8\%\\
AutoVCoder-Codellama-ER\&KR & \textbf{63.7\%} & \textbf{72.9\%} & 43.2\% & \textbf{52.8\%} & \textbf{93.1\%} & \textbf{48.3\%} \\

\hline
AutoVCoder-DeepSeek-1\&2 & 67.1\% & 77.8\% & 45.1\% & 52.8\% & \textbf{100\%} & \textbf{51.7\%} \\
AutoVCoder-DeepSeek-ER & 68.1\% & 79.1\% & 46.6\% & 53.1\% & \textbf{100\%} & \textbf{51.7\%}\\
AutoVCoder-DeepSeek-KR & 66.9\% & 77.5\% & 45.5\% & 53.3\% & \textbf{100\%} & \textbf{51.7\%}\\
AutoVCoder-DeepSeek-ER\&KR & \textbf{69.0\%} & \textbf{79.3\%} & \textbf{46.9\%} & \textbf{53.7\%} & \textbf{100\%}  & \textbf{51.7\%} \\

\hline
AutoVCoder-CodeQwen-1\&2 & 66.8\% & 78.3\% & 46.2\% & 54.1\% & \textbf{100\%} & \textbf{51.7\%}\\
AutoVCoder-CodeQwen-ER & 68.2\% & 79.1\% & \textbf{48.5\%} & 55.3\% & \textbf{100\%} & \textbf{51.7\%}\\
AutoVCoder-CodeQwen-KR & 68.5\% & 79.3\% & 46.5\% & 54.0\% & \textbf{100\%} & \textbf{51.7\%}\\
AutoVCoder-CodeQwen-ER\&KR & \textbf{68.7\%} & \textbf{79.9\%} & 48.3\% & \textbf{55.9\%} & \textbf{100\%} & \textbf{51.7\%}\\

\hline
\end{tabular}
\vspace{-0.37cm}
\label{tab_3}
\end{table*}

\subsection{Experimental Settings}
\vspace{-1mm}
%In our fine-tuning process, we select the well-trained models for our base models. Our initial base models include Codellama-7B~\cite{rozière2024code}, Mistral-7B-v0.1~\cite{jiang2023mistral}, and Starling-LM-7B-alpha\cite{starling2023}. For two rounds of model fine-tuning, we applied the LoRA method for parameter-efficient fine-tuning across all models. We set the learning rate at $\gamma = 2e-4$. We carry out our fine-tuning tasks on three Nvidia A100 GPUs. We train 1 epoch for the first-round fine-tuning and 3 epoches for the second-round fine-tuning. For RAG retriever training, we utilized FlagEmbedding for the base sentence embeddings model and adopt a method of full parameter fine-tuning. We set the learning rate at $\gamma = 1e-5$ and train it for 3 epoches. This training task is also performed on three Nvidia A100 GPUs.

We use some open-source pre-trained LLMs as our base models, including Codellama-7B~\cite{rozière2024code}, DeepSeek-Coder-6.7B~\cite{guo2024deepseekcoderlargelanguagemodel}, and 
CodeQwen1.5-7B~\cite{bai2023qwentechnicalreport}. During the model fine-tuning, we use the LoRA~\cite{hu2022lora} method to maintain high efficiency, and we set the learning rate $\gamma$ as $2e-4$. We train the LLMs for one epoch in the first-round fine-tuning and three epoches in the second-round fine-tuning. As for RAG retriever training, we utilize FlagEmbedding~\cite{llm_embedder} to extract base sentence embeddings, which is trained for three epoches at a learning rate of $1e-5$. The training and inference processes are carried out on three Nvidia A100 GPUs.

To evaluate the models in the inference stage, two key hyperparameters related to LLM, i.e., $top_p$ and $temperature$, are set as $0.95$ and $0.8$, respectively. Moreover, we adopt a widely used evaluation metric \emph{pass@$k$}~\cite{chen2021evaluating} for code generation, which refers to the probability that a code solution passes validation when generated $k$ times. This metric can be calculated as:
\[
\text{pass@}k = \mathbb{E} \left(1 - \frac{{\binom{n-c}{k}}}{{\binom{n}{k}}}\right),
\]
where $n$ is the total number of test for the task and $c$ is the number of correct code generations for the task. We set $n$ as $10$ in our experiments and use the pass@1 metric and pass@5 metric for evaluation. 

To evaluate the effectiveness of our Verilog code generation, we utilized two up-to-date RTL benchmarks: VerilogEval~\cite{liu2023verilogeval} and RTLLM V1.1~\cite{liu2024rtlcoder}. VerilogEval is divided into two sub-tasks, i.e., EvalMachine and EvalHuman, each of which comprises over 100 RTL design tasks. EvalMachine contains Verilog questions that are generated automatically, while those of EvalHuman are manually designed. RTLLM V1.1 includes 29 RTL design tasks. We evaluate the design effectiveness from two aspects: syntactic correctness and functional correctness. Accordingly, we utilize Icarus Verilog~\cite{aslan2016open} and PyVerilog to check the syntax correctness of the generated Verilog solutions. If the designs are syntactically correct, we then run the testbenches from the benchmarks, and compare the results of the generated solutions with the golden output to examine their functional correctness.

\vspace{-0.1cm}
\subsection{End-to-end Comparison with SOTA Methods}
% \subsection{Syntactic and Functional Correctness}
In this experiment, we evaluate the syntactic and functional correctness of the LLMs augmented with our proposed two-round fine-tuning and domain-specific RAG techniques, and finally compare the results with SOTA methods from both the industry and academia. Experimental results are demonstrated in Table \ref{tab_2}, where the best and second-best results are marked in green and yellow, respectively. Among the large-scale LLMs from the industry, GPT-4 outperforms GPT-3.5 and achieves superior results in both the VerilogEval and RTLLM question sets. Among small-scale models with less than 16B parameters, our proposed method, AutoVCoder, presents the best overall performance in most cases and achieves top-2 correctness in all cases, outperforming the SOTA research works. It shows a 0.5\% and 2.2\% improvement in functional correctness on the EvalMachine and EvalHuman benchmarks compared with BetterV, and achieves an increase in syntactic and functional correctness of up to 3.4\% on the RTLLM benchmark compared with RTLCoder. In summary, when compared with the SOTA method using small-scale LLMs, AutoVCoder performs the best, especially for the two realistic question sets, EvalHuman and RTLLM. Moreover, it is worth mentioning that AutoVCoder also outperforms ChatGPT-4, the representative large-scale LLM, regarding the EvalMachine question set. The results verify the effectiveness of the strategies proposed by AutoVCoder, including the dataset generation, fine-tuning and domain-specific RAG.

% \vspace{-2mm}
\subsection{Improvement from Two-Round Fine-Tuning}
%In this experiment, we fine-tune the base models through two rounds as described in our methodology. To validate the effectiveness of our two-round fine-tuning framework, we also test the models which skip the first round  and directly fine-tuned in the second round. The comparative analysis focused on syntax accuracy and functional accuracy. Models with \verb|-1| suffix indicate those models fine-tuned in the first round, while models with \verb|-2| suffix skip the first round and are only fine-tuned in the second round. Models with a \verb|-1&2| suffix undergo complete two rounds of fine-tuning.

In this experiment, we evaluate the benefits of applying our two-round fine-tuning method. To quantify the efficacy of each round of fine-tuning, we conduct experiments with the following four settings: (1) no fine-tuning; (2) only applying the first-round fine-tuning, denoted with the suffix \verb|-1|; (3) only using the second-round fine-tuning, denoted with the suffix \verb|-2|; and (4) employing both the two rounds of fine-tuning, denoted with the suffix \verb|-1&2|. 

Table \ref{tab_1} illustrates the experimental results, which shows that both the first-round and second-round fine-tuning are crucial in boosting the syntactic and functional correctness, compared with the base models. Comparatively speaking, the second-round fine-tuning exerts a more significant impact on the correctness improvement over the first-round fine-tuning. This showcases the benefits of the high-quality and highly specialized dataset construction method for the second-round fine-tuning. However, the first-round fine-tuning is also indispensable in letting LLMs learn from a more diversified database, which ensures high generalization ability of LLMs. 

In terms of the base models used, CodeQwen1.5-7B performs the best, which achieves 78.3\% and 54.1\% functional correctness on EvalMachine and EvalHuman, respectively, and achieve 100\% syntax accuracy and 51.7\% functional accuracy on the RTLLM benchmark. This also confirms that the LLMs are promising in the task of automatic Verilog code generation by integrating appropriate fine-tuning techniques.

%\begin{table}[t]
%\centering
%\caption{Comparison between AutoVCoder and AutoVCoder with RAG Technique}

%\scalebox{0.85}{
%\begin{tabular}{|c|c|c|c|c|}
%\hline
%\multirow{2}{*}{Evaluated Model} & \multicolumn{2}{c|}{\makecell{VerilogEval \\ (pass@5)}} & \multicolumn{2}{c|}{\makecell{RTLLM V1.1 \\ (pass@5)}}   \\ \cline{2-5} 
                                 %& \multicolumn{1}{c|}{EvalMachine}  & EvalHuman  & %\multicolumn{1}{c|}{Syn.} & Func. \\ \hline

%AutoVCoder-Codellama-1\&2 & 66.6\% & 47.1\% & 89.6\%  & 44.8\% \\
%AutoVCoder-Codellama-1\&2$^G$ & 72.9\% & 52.8\% & 93.1\%  & 48.3\% \\ \hline
%AutoVCoder-Mistral-1\&2 & 69.1\% & 45.0\% & 96.6\%  & 51.7\% \\
%AutoVCoder-Mistral-1\&2$^G$ & 72.3\% & 48.1\% & 96.6\%  & 51.7\% \\ \hline
%AutoVCoder-Starling-1\&2 & 70.9\% & 51.1\% & 100\%  & 55.2\% \\
%AutoVCoder-Starling-1\&2$^G$ & 73.9\% & 52.2\% & 100\%  & 55.2\% \\

%\hline

%\end{tabular}
%}

%\label{tab_3}

%\end{table}

%\vspace{-2mm}
\subsection{Improvement from Domain-Specific RAG}
%To validate the effectiveness of our RAG module, we deploy the RAG module to our three two-round fine-tuned LLM and compare their performance against their original model without the RAG enhancement. In this experiment, we used the RAG module to retrieve two examples and three knowledge chunks. We tested the pass@5 metric on two benchmarks. The results are shown in the Table \ref{tab_3}, with the superscript G indicating models with the RAG module. The results showed a notable improvement on VerilogEval. After deploying the RAG module, the three models show functionality accuracy improvements on EvalMachine of 6.3\%, 3.2\%, and 3\%, respectively, and on EvalHuman of 5.7\%, 3.1\%, and 1.1\%, respectively. We find that the accuracy improve the most in the Codellama model, likely because our retriever's dataset is generated based on this model. The improvements on RTLLM are less significant, only Codellama model gains improvement, suggesting that our collected knowledge and example documents are not well-matched to some of the difficult problems in RTLLM. We will continue to optimize the documents in future work. In summary, our RAG module can enhance model accuracy on benchmarks to some extent and represents a step forward in expanding our models to new RTL domains.

In this experiment, we verify the effectiveness of the domain-specific RAG technique. We separately examine the following cases given the models after the two-round fine-tuning: (1) not applying RAG; (2) only using the example retriever, denoted as ER; (3) only using the knowledge retriever, denoted as KR; and (4) applying both the example retriever and knowledge retriever, denoted as ER\&KR. We evaluate the pass@5 metric on VerilogEval and RTLLM question sets and the results are shown in the Table \ref{tab_3}. 

Results indicate that the example retriever is prominent in enhancing the LLM's capability of Verilog coding, especially for the VerilogEval question set. Moreover, even though the standalone employment of the knowledge retriever does not make significant gain, jointly using both the example retriever and the knowledge retriever achieves the best performance in most cases, which is a step forward in expanding the LLMs to new RTL domains. In addition, we observe that the improvements on RTLLM question set are less notable. The main reason is that the example database is not well-matched to some of the difficult problems in RTLLM, due to the scarcity of large-scale Verilog design instances. We believe that this problem can be alleviated by continuously accumulating large-scale and realistic design instances and keep expanding the example database.

% \vspace{-2mm}
\section{Conclusion}

In conclusion, we propose AutoVCoder, a systematic framework for Verilog code generation using LLMs. We introduce three innovative methods to enhance the LLMs' capability in generating high-quality Verilog code, which effectively improves the level of automation for digital circuits. Experimental results demonstrate that AutoVCoder outperforms existing state-of-the-art methods, paving the way for efficient and effective hardware design using natural languages.
% Firstly, we propose an automated dataset generation approach that provides high-quality and diverse RTL code samples with few human intervention. Secondly, we introduce a two-round fine-tuning process for LLM, improving the LLMs' efficiency and quality of results. Thirdly, we present a domain-specific RAG technique that provides more constructive prompts to further enhance the syntactic and functional correctness of the generated hardware designs. Putting it all together, this paper paves the way for efficient and effective hardware design using natural languages.

% \vspace{-0.3cm} 

\bibliographystyle{IEEEtran}
\bibliography{ref}

% \end{thebibliography}
\end{document}